\documentclass{article}
\usepackage{rapatt}
\usepackage{epsfig}
\usepackage{graphicx}
\usepackage{cancel}
\usepackage{amsmath,amssymb}
\usepackage{color}
\usepackage[hyphens]{url}
\begin{document}

\centerline{{\large{\bf DA$\Phi$NE - 2023/24 Activity report}}\footnote{Published in 2023 LNF Activity report: \url{http://www.lnf.infn.it/rapatt/2023/dafne.pdf}} 
}
\begin{center}
C. Milardi\footnote{Scientific head. Mail: catia.milardi@lnf.infn.it}, D. Alesini, M. Behtouei,  S. Bilanishvili, S. Bini, M. Boscolo, B. Buonomo, \\ S. Cantarella, A. Ciarma, A. De Santis\footnote{Author: antonio.desantis@lnf.infn.it}, E. Di Pasquale, C. Di Giulio, G. Di Pirro,\\ 
O. Etisken, L. Foggetta, G. Franzini, A. Gallo, R. Gargana, S. Incremona, A. Liedl, A. Michelotti, \\ L. Piersanti, D. Quartullo, R. Ricci, U. Rotundo, S. Spampinati, A. Stecchi, \\ A. Stella,  A. Vannozzi, M. Zobov.

\end{center}

\section{Introduction}

The DA$\Phi$NE accelerator complex~\cite{dafne} is a double-ring lepton collider operating at a center-of-mass energy of 1.02 GeV ($\Phi$-resonance) with a full-energy injection system. The infrastructure consists of two independent rings (Main Rings - MRs), each 97 meters long, with intersection points at the Interaction Region (IR) and Ring Crossing Region (RCR), where beams cross at a horizontal angle of 50 mrad. DA$\Phi$NE also provides synchrotron light lines~\cite{dafneLight} and a Beam Test Facility~\cite{BTF}. Despite being constructed in the 1990s, DA$\Phi$NE remains a unique machine for studying low-energy kaons with momenta below 140 MeV/c. It pioneered the Crab-Waist collision scheme~\cite{crab1,CWSiddharta}, successfully boosting luminosity by a factor of three and becoming the benchmark approach for modern and future lepton colliders\cite{fcwc1,fcwc2,fcwc3,fcwc4,fcwc5,fcwc6}. 
\par
Since 2021, DA$\Phi$NE has been providing data to the SIDDHARTA experiment~\cite{runSiddharta21}, enabling high-precision kaonic helium measurements, first with the preliminary SIDDHARTINO~\cite{helium4perf,helium4meas} setup and later with the final SIDDHARTA-2 configuration.

\section{Last year run}

The DA$\Phi$NE operations during the last year \cite{Milardi:ipac2024-wepr17} have been devoted to deliver a statistically significant data sample to perform the first-ever measurement of kaonic deuterium X-ray transitions to the fundamental level \cite{deuteriumMeas}.
Operations for the SIDDHARTA-2 detector using a deuterium gas target started officially on the second half of May 2023, and have been organized in several runs.
Initially the efforts on the DA$\Phi$NE side were aimed at optimizing injection and collisions setup, while SIDDHARTA-2 group concentrated on testing the new experimental apparatus, and tuning the setup to maximize the signal to noise ratio. Thereafter, operations continued privileging data delivery with short periods dedicated to machine studies. After completing operations with the deuterium target, a calibration run with hydrogen gas followed, intended for further detector characterization and background studies.

\section{Collider tuning}

During the final operational period, the maximum injection efficiency for both beams reached 80\%, with transport efficiency along the transfer lines close to 100\%. These optimal efficiencies played a significant role in enabling the storage of high beam currents and fine-tuning the collision process. These improvements were critical in supporting data delivery with stable beam conditions.

\subsection{Linear and Non-Linear Optics}

The optics used in MRs were optimized in previous runs~\cite{DAFNE_IPAC23}, which had undergone continuous refinement. The Crab-Waist sextupoles strengths had been gradually increased to approximately 77\% of their nominal values. This adjustment led to a significant improvement in instantaneous luminosity and enhanced the control over background noise.

A detailed optimization process of the chromatic sextupoles and octupoles, conducted iteratively during data collection, further improved beam lifetime and injection efficiency while reducing the signal-to-noise ratio. As a result, the background noise affecting detector measurements was greatly reduced, contributing to cleaner data and more accurate results.

\begin{figure}[!h]
    \centering
    \includegraphics[width=0.57\linewidth]{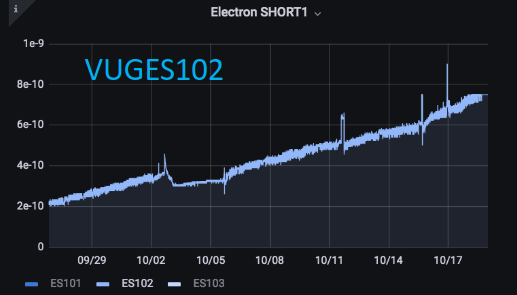}
    \includegraphics[width=0.37\linewidth]{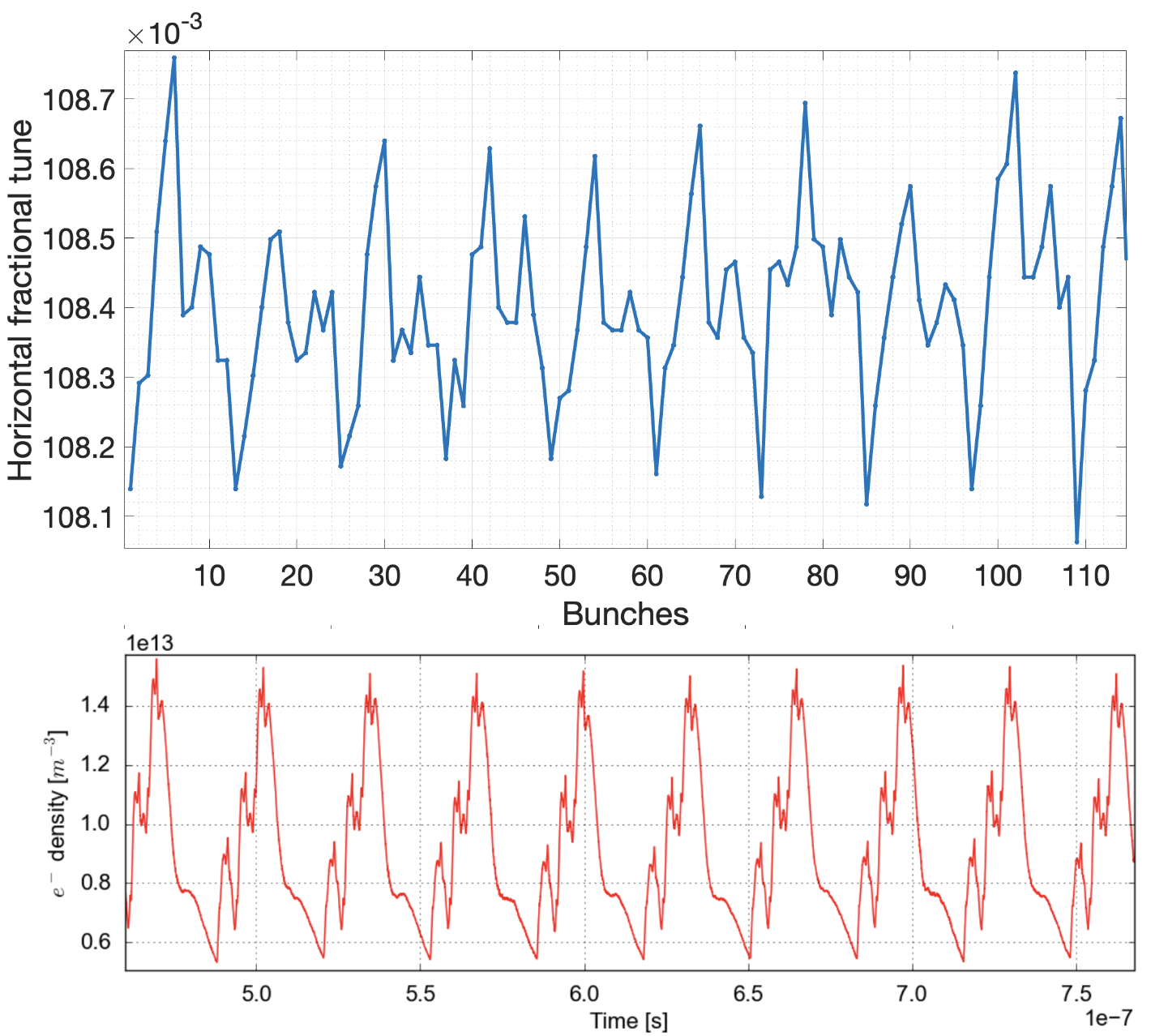}
    \caption{Left: vacuum pressure rise in the MRe before the leak was found. Right: Horizontal tune shift measurement for 60 bunches (each train has 6 filled 6 empty buckets) which correspond to 290 mA beam current (top) and e-cloud build-up simulation by PyECLOUD (bottom).}
    \label{beamDyn}
\end{figure}

\subsection{Beam Dynamics}
Since fall 2023 several anomalous beam dynamics trends, particularly affecting the electron beam, were observed. The vertical tune has been lowered from its nominal value to ensure beam stability while a reduced beam lifetime, and strong vertical instabilities at relatively low current levels appeared. Sudden beam losses also occurred for electron beam intensities above 1.5 A. Collision operations, despite the beam-beam stabilizing effect, were impacted by the flip-flop effect. The electron beam experienced also vertical blow-up, leading to a sharp drop in luminosity and an increase in background noise. These symptoms were attributed to vacuum condition worsening (as shown in Fig.~\ref{beamDyn}-left) leading to ion trapping-induced instabilities. To mitigate these issues, the number of bunches in collision was reduced from 110 to 95. This temporary adjustment allowed for continued data collection while investigations were underway. A vacuum leak inspection later revealed a small leak (on the order of $10^{-8}$ mbar) in the first short arc of MRe. Once the leak was repaired, the electron beam dynamics stabilized, and collisions with 110 bunches were restored.

Throughout the collider operations, e-cloud effects posed significant challenges for positron beam dynamics. These effects were mitigated through a combination of strategies, including solenoid windings around the beam pipes, transverse feedback systems, tuning positron beam parameters, reducing RF cavity voltage to lengthen bunches, and adding Landau damping using octupole magnets. Given the importance of e-cloud instabilities, a campaign of simulations and measurements had been launched. These studies aimed not only to address the immediate challenges at DA$\Phi$NE but also to provide valuable insights for future collider projects, such as FCC-ee~\cite{fcwc2} helping to benchmark existing numerical codes~\cite{ecloud}.
The density of the electron cloud induced by the positron beam in non-colliding configurations was measured at currents up to 800 mA. The horizontal tune shifts along the bunch train were used to calculate the electron cloud density, which was found to be as high as $10^{14} \mbox{cm}^{-3}$, one of the highest recorded in any circular collider (see Fig.~\ref{beamDyn}-right). This high density posed significant challenges to beam stability.
\par
Grow-damp measurements were performed to study the multi-bunch horizontal instability in the positron beam due to e-cloud effects~\cite{quartullo:ibic24}. By deliberately turning off the horizontal feedback system for a specific time interval, the instability was induced, and the bunch-position signals were analyzed. The dominant instability mode, m = -1, was observed with a growth rate of 22 ms$^{-1}$. This mode is due to e-cloud created in wigglers and bending magnets and it is consistent with earlier observations from a 2008 analysis~\cite{drago}.
\par
During the final phase of operations, DA$\Phi$NE achieved significant improvements in beam dynamics, luminosity, and background noise control through a combination of optical adjustments and the mitigation of e-cloud and ion trapping instabilities. The collider reached stable beam currents of up to 1.0 A for positrons and 1.65 A for electrons during collisions. Additionally, increasing the vertical chromaticity to +1.5 proved effective in mitigating e-cloud effects on the positron beam.

\section{Collider performances}
\subsection{Luminosity}
Luminosity at DA$\Phi$NE was monitored using two main devices: Crystal CALorimeters (CCAL)~~\cite{ccallumikloe} and Gamma monitors. CCAL detected Bhabha scattering events at small angles and, due to its high event rate, was used as a real-time tool for luminosity optimization and was extremely effective also to monitor beam induced background rate and spatial distribution. However, the CCAL was not fully calibrated, and the Gamma monitor suffered from beam losses, preventing them from providing reliable absolute luminosity measurements.
\par
Absolute luminosity was instead measured by the SIDDHARTA-2 detector, which detects charged kaon flux~\cite{lumisid}. In addition, real-time background levels in the experimental apparatus were monitored by counters measuring Kaon over Minimum Ionizing Particle rate (Kaon/MIP) and Kaon over Silicon Drift Detector rate (Kaon/SDD). The Kaon/SDD rate was also used as a key quality parameter (L$_{\text{HQ}}$), indicating whether data could be used for physics analysis.
\par
\begin{figure}[!h]
    \centering
    \includegraphics[width=0.45\linewidth]{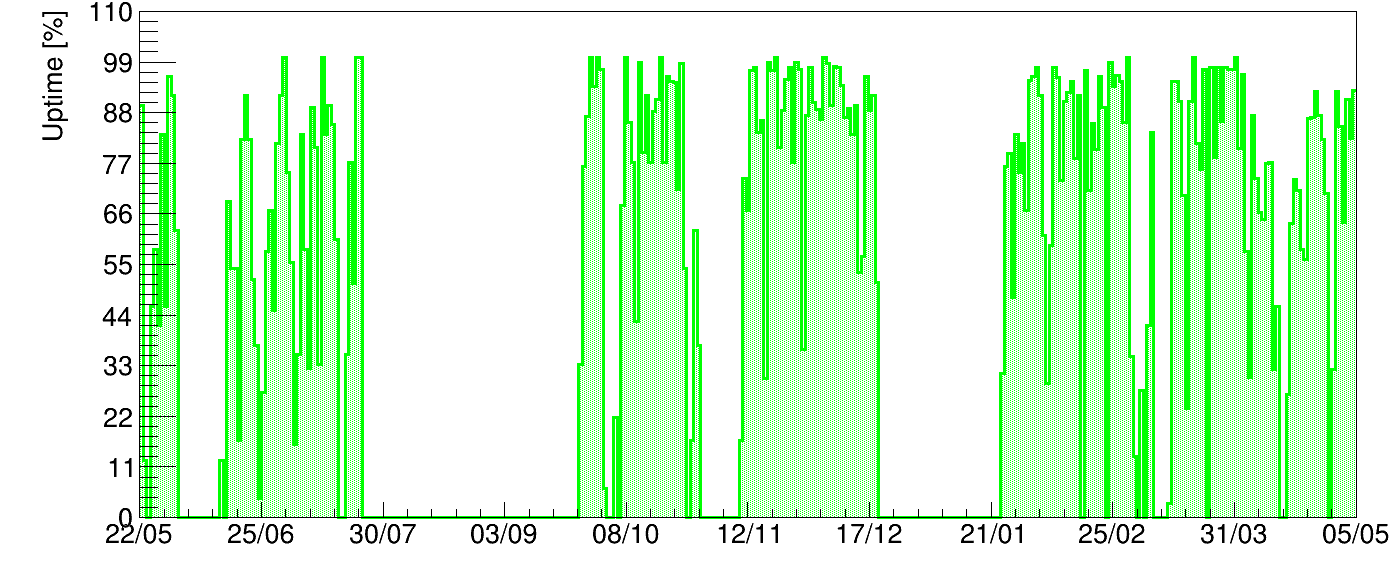}
    \includegraphics[width=0.53\linewidth]{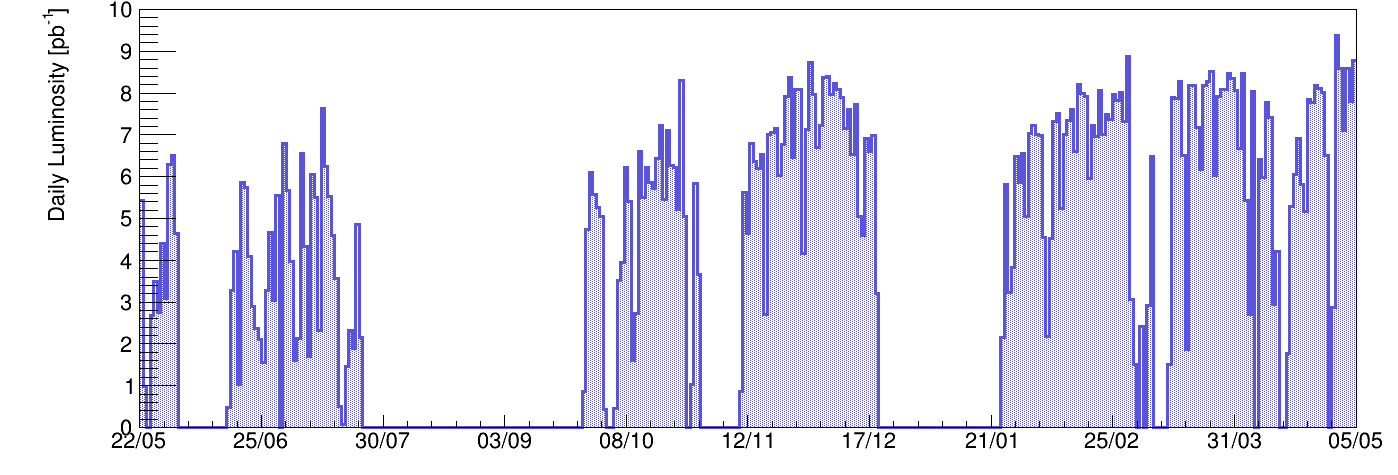}
    \caption{Left: Luminosity delivering efficiency. Excluding the planned operation interruptions during holidays, the average value for the whole period is around 75\%.
    Right: Daily delivered luminosity. A maximum value of 9.5 pb$^{-1}$ is observed.}
    \label{fig:lumideliv}
\end{figure}

The highest instantaneous luminosity recorded was $2.4\times10^{32}\mbox{cm}^{-2}\mbox{s}^{-1}$, achieved with an electron beam current of 1.14 A and a positron beam current of 0.89 A, both stored in 110 bunches. The average efficiency of the collider, defined as the percentage of time it delivered luminosity above $10^{32}\mbox{cm}^{-2}\mbox{s}^{-1}$ after beam refilling, was around 75\%, as shown in Fig.~\ref{fig:lumideliv}-left.
\par

Although downtime including maintenance, technical faults, and institutional duties, the overall trend for daily delivered luminosity remained positive. The maximum daily luminosity recorded reached approximately 9.5 pb$^{-1}$, as measured by the kaon monitor reported in Fig.~\ref{fig:lumideliv}-right.

\par
\begin{figure}[!h]
    \centering
    \includegraphics[width=0.8\linewidth]{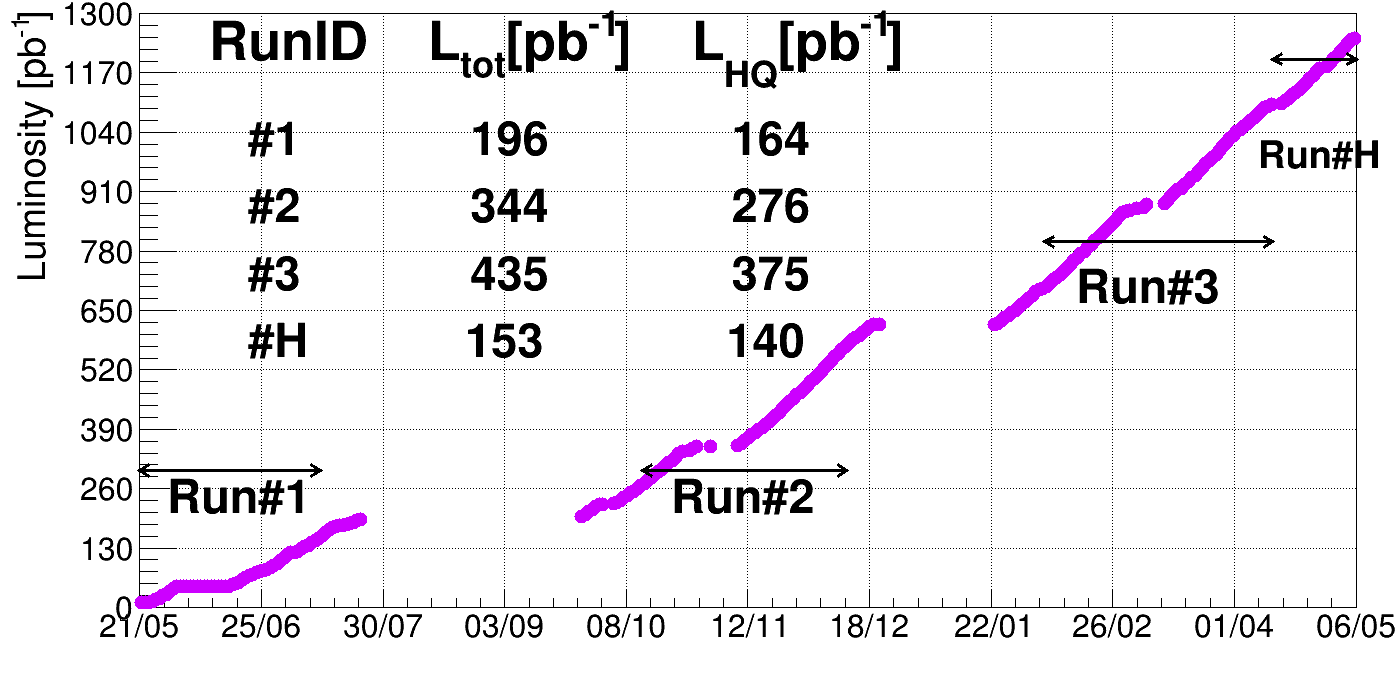}
    \caption{Total acquired luminosity. In the table the total delivered (L$_{\text{tot}}$) and the high quality (L$_{\text{HQ}}$) integrated luminosity is reported for each run period. The fraction of high-quality data increased significantly along the time as a result of the tuning of the DA$\Phi$NE collider.}
    \label{fig:totalDelLumi}
\end{figure}

The last year operations have been splitted in three operational runs, each of similar duration and uptime. The total delivered luminosity exhibits a substantial increase, as demonstrated in~Fig.~\ref{fig:totalDelLumi}. In the Run~3 operations, luminosity delivery more than doubled compared to Run~1, marking a significant improvement in the collider’s performance.

\subsection{Background}

DA$\Phi$NE achieved significant progress in background reduction during its final operations, particularly in addressing two types of background: injection and coasting background.

\begin{description}
    \item[Injection Background:] Injection-related background was reduced through the optimization of injection efficiency and precise steering of the stored beam orbit. This approach successfully reduced the background for the electron beam but was less effective for the positron beam. To address this, the vertical size of the electron beam was artificially increased during injection using a calibrated skew quadrupole bump, thereby reducing the beam-beam interaction on the positron beam and preventing rapid lifetime drops and sudden background surges.
    \item[Coasting Background:] Coasting background was minimized by reducing Kaon/MIP and Kaon/SDD rates through a comprehensive optimization of the collider’s non-linear optics. The tuning of sextupole and octupole magnets in both rings increased energy acceptance and dynamic aperture, leading to a nearly twofold improvement in Kaon/MIP rates and a 1.45-fold improvement in Kaon/SDD rates.
\end{description}
    
The improvements in background reduction resulted in a significantly enhanced signal-to-noise ratio (SNR) for the SIDDHARTA-2 experiment compared to the 2009 Crab-Waist test run \cite{bckComp_09_24}. The SNR was three times higher than in the 2009 run, thanks to the optimization of the collider configuration and upgrades to detector components, such as kaon, trigger, and Silicon Drift Detector (SDD) systems. Preliminary analysis indicated that these gains were largely due to collider optimization and the new design of the Permanent Magnet Defocusing Quadrupole installed in the low-beta section of the interaction region.

\section{Conclusion}

In its final operational phase, DA$\Phi$NE demonstrated considerable improvements in luminosity delivery and background reduction. The peak luminosity of $2.4\times10^{32}\mbox{cm}^{-2}\mbox{s}^{-1}$ and the consistent trend toward higher daily luminosity reflect the success of targeted optimizations. The collider operated with an efficiency of around 75\%, delivering reliable performance despite routine maintenance and downtime. Moreover, background reduction strategies—particularly for injection and coasting phases—led to enhanced data quality, as evidenced by the improved signal-to-noise ratio. The DA$\Phi$NE lepton collider has delivered to the SIDDHARTA-2 detector using a deuterium gas target a data sample of the order of 1.24 fb$^{-1}$, well beyond the experiment request. 
\par
These achievements represent the culmination of extensive tuning efforts and the lessons learned from these developments, particularly in the areas of Crab-Waist sextupole tuning and instability mitigation, have provided valuable insights for future high-luminosity colliders such as FCC-ee. 

\section{ACKNOWLEDGEMENTS}
Many thanks to the Staff of the Accelerator and Technical Divisions of the LNF. Special acknowledgments to the operation group taking care of the collider operations 24 hours a day, their commitment largely contributed to achieve the present DA$\Phi$NE performances.

%
% only for "biblatex"
%

\end{document}